\documentstyle[twoside,fleqn,espcrvj,epsfig,subfigure]{article}
\newcommand{\Ds}{\mbox{D}_s}

\newcommand{\Dstn}{\mbox{D}^{\ast 0}}
\newcommand{\Dstp}{\mbox{D}^{\ast +}}

\newcommand{\Dsts}{\mbox{D}^{\ast}_s}

\newcommand{\ccs}{b\rightarrow c{\bar c}s}
\newcommand{\cud}{b\rightarrow c{\bar u}d}

\newcommand{\AmS}{{\protect\the\textfont2
  A\kern-.1667em\lower.5ex\hbox{M}\kern-.125emS}}

\title{Recent results from CLEO on Charm and Bottom hadrons}

\author{Vivek Jain\address {Vanderbilt University,\\  
Nashville, TN 37235, USA \\
Representing the CLEO collaboration}}

\begin{document}

\begin{abstract}
In this talk\footnotemark, I present new results from CLEO on charm and bottom hadrons. Most of the talk will be
 on the 
issue of the B semileptonic branching fraction, its connection to the number of charm quarks 
produced in the decay of a b quark, and the rate for the $\ccs$ transition.
 \end{abstract}

\maketitle

\section{Introduction}
\footnotetext{Invited talk at ``Production 
and decay of hyperons, charmed and beauty hadrons'', Strasbourg, France, Sep. 5-8, 1995}
   The physics program at CLEO is at the forefront of heavy flavour research. The emphasis is on the
decay of charm hadrons, beauty mesons and tau leptons. There is also active research in 
2-photon physics, Upsilon spectroscopy and production characteristics of charm hadrons. 

   In this talk, I will focus on the disagreement between the experimental value of the 
B semileptonic branching fraction and predictions of theoretical models; the experimental value being 
the smaller of the two.
In order to ``fix'' the model predictions, one has to increase the number of charm quarks produced in 
the decay of a b quark, and also the rate for B decays of the type, $\ccs$. I will discuss CLEO results
which shed light on this issue. I will first present results on an isospin violating decay of 
the $\Dsts$ meson.

\section{Data Sample}

     The results shown here are based on data taken at the Cornell Electron Storage Ring using the
CLEO-II detector. The CLEO-II detector has excellent charged and neutral particle detection over 
$\approx 95\%$ of $4\pi$. Electrons and muons are detected with high efficiency and low fake rates. 
Detector details can be found elsewhere\cite{ref:nim}. 

     The data were collected on the $\Upsilon$(4S) resonance, with center of mass energy
of 10.58 GeV, and in the continuum, 60 MeV below. The ON resonance luminosity was 3.3 fb$^{-1}$, 
which corresponds to about $3.5 \times 10^6$ $B {\bar B}$ mesons produced. The OFF resonance luminosity,
 which is used to model the continuum background under the $\Upsilon$(4S), 
was 1.6 fb$^{-1}$. To study charm hadrons, one can use both ON and OFF resonance data, which amounts to
about $6.5~\times~10^6$ $c {\bar c}$ pairs produced. The total number of reconstructed charm hadrons 
at present, which includes $D^0, D^+, D^{*0(+)}, D_s^{(*)}, \Lambda_c$, etc., is $\ge~1.0\times~10^5$.
The results presented here are based on about 70\% of the total luminosity.

\section{Isospin violating decay, $\Dsts \rightarrow \Ds \pi^0$}

    Up to now, only the radiative decay of the $\Dsts$ has been observed\cite{ref:pdg}.  The only strong 
decay allowed, $\Dsts \rightarrow \Ds \pi^0$, is ``forbidden'' by isospin. However, isospin is not an
exact symmetry, e.g., $m_u \ne m_d$, and the presence of the decay 
$\psi^{'}~\rightarrow~J/ \psi \pi^0$. It has
been argued on the basis of chiral perturbation theory that $\Dsts \rightarrow \Ds \pi^0$ is 
non-vanishing.
The decay is mediated by a virtual $\eta$, which has a significant $s {\bar s}$ content, which then 
``mixes'' into a $\pi^0$, due to the fact that the former also has a large non-strange component. The 
second step violates isospin. The tree level diagram for this decay, gluon emission to produce a 
$\pi^0$, is OZI-suppressed, whereas the electromagnetic production mechanism is down by a factor of 
$\alpha$. The amplitude for this decay mode is proportional to the mass difference 
between the u and d quarks. Since the radiative decay, $\Dsts \rightarrow \Ds \gamma$, is suppressed 
due to the partial cancellation of the charm and strange quark magnetic moments, it is possible to 
observe the isospin violating decay.

\begin{figure}[htb]
\vspace{9pt}
\hspace{0mm}
\vspace{-0.5cm}
{\epsfig{figure=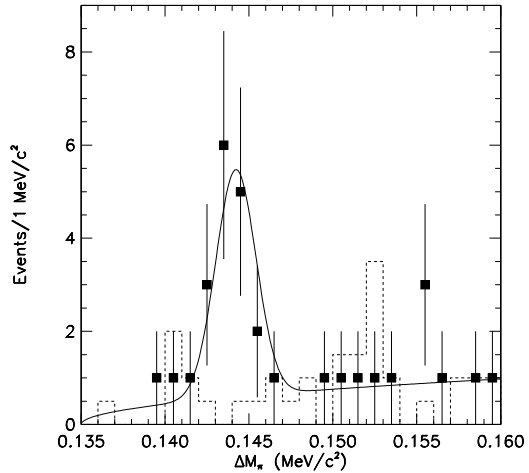, height=6.43 cm}} 
\vspace{-0.5cm}
\caption{ Mass Difference.}
\label{fig:pizero}
\end{figure}

    The $\Ds$ meson is reconstructed in the $\phi \pi$ decay mode, which has a large 
(detection efficiency $\times$ branching fraction) and is relatively background free\cite{ref:bart}. 
The $\pi^0$ has to pass strict selection criteria in order
to be considered. In Fig.~\ref{fig:pizero}, I present the mass difference, $\Delta M = M(\Ds \pi^0) - M(\Ds)$, 
for events which fall within the $\pi^0$ and $\Ds$ mass regions. The points with error bars 
indicate a clear
signal, yielding $14.7^{+4.6}_{-4.0}$ events. The dashed line is the contribution due to 
random combinations,
which has been modelled using the sidebands in the $\pi^0$ and $\Ds$ mass distributions. A fit to the 
dashed histogram yields $-1.0^{+3.1}_{-2.4}$ events, consistent with zero. If, instead, we
plot the $\Ds$ mass, after requiring cuts on the mass difference, we again have a clear signal.
Counting events in the signal region, 142~MeV/c$^2~<~\Delta M~<~146$~MeV/c$^2$, 
we observe 16 signal and 5
background events. Taking into account that the sidebands are twice the width of the signal region, we
obtain 
the binomial probability of getting 16 (or more) signal events out of a total of 21 events to be 
$7.3 \times 10^{-5}$, which corresponds to a statistical significance of at least 3.9 standard 
deviations. Normalizing this reaction to the radiative decay, we obtain the branching fraction ratio,

\begin{displaymath}
\frac{{\cal B} (\Dsts \rightarrow \Ds \pi^0)}{{\cal B} 
(\Dsts \rightarrow \Ds \gamma)}~=0.062^{+0.020}_{-0.018} \pm 0.022 
\end{displaymath}

    The presence of both the radiative and pionic decay modes implies that
the spin-parity of the $\Dsts$ belongs to the ``natural'' series ($1^-$, 2$^+$,...). The most likely
scenario is $1^-$, same as $\Dstn$ and $\Dstp$ \cite{ref:pdg}. In addition, the pionic 
decay mode
is very close to the kinematic threshold; we use it to measure the  mass difference of 
$\Dsts$ and $\Ds$, which is determined to be $143.76 \pm 0.39 \pm 0.40$ MeV/c$^2$, in excellent 
agreement with the
previous CLEO measurement (using the radiative mode), $144.22 \pm 0.47 \pm 0.37$ MeV/c$^2$. 
These values are somewhat larger but more precise than the 
PDG\cite{ref:pdg} value of $142.4 \pm 1.7$ MeV/c$^2$.

\section{Semileptonic B decay and related issues}

    One of the more intriguing issues in B physics is the disagreement between the experimental value 
and theoretical predictions for the B semileptonic branching fraction. After accounting for QCD 
corrections, the theoretical predictions range from $11\% - 12\%$, whereas the most model independent
experimental value (CLEO) is $(10.49 \pm 0.17 \pm 0.43)\%$. 
This ``disagreement'' may not seem real, but the 
problem is that these theoretical models also predict that the number of charm quarks ($n_c$) 
produced per decay
of a b quark is about 1.30 instead of the measured value (CLEO),
\begin{displaymath}
 n_c = 1.15 \pm 0.044 
\end{displaymath}
These predictions imply 
that the rate of the $\ccs$ transition is boosted from 0.15 to about 0.30; the 
lower the theoretical prediction for ${\cal B} (B \rightarrow X l \nu)$, the higher the prediction for
$n_c$ and $\Gamma (\ccs)$.
Table~\ref{tab:nc} lists the latest CLEO results on the inclusive decay rates of the B meson 
into various charm final states\cite{ref:note}.

\begin{table}[hbt]
\setlength{\tabcolsep}{2.0pc}
\caption{Inclusive B decays to charm hadrons.}
\label{tab:nc}
\begin{tabular}{cc}
\hline
  Decay mode &  Rate \\ 
\hline
 ${\bar B} \rightarrow \mbox{D}^0 X$ &  $(64.6 \pm 3.2)\% $      \\
 ${\bar B} \rightarrow \mbox{D}^+ X$ &  $(25.3 \pm 1.6)\% $  \\
 ${\bar B} \rightarrow \Ds^+ X$ & $(11.8 \pm 1.7)\% $  \\
 ${\bar B} \rightarrow \Lambda_c X$ & $(4.0 \pm 1.0)\% $  \\
 ${\bar B} \rightarrow \Xi_c X$ & $(3.9 \pm 1.8)\% $  \\
 ${\bar B} \rightarrow c{\bar c} X$ & $(5.2 \pm 0.7)\% $  \\
\hline
 $n_c$ & $1.15 \pm 0.044$          \\
\hline
\end{tabular}
\end{table}


    If these theoretical models are right then $\Gamma (\ccs) \approx 0.30$, and 
$\Gamma(\ccs)/\Gamma(\cud)~\approx~2/3$.
This does not change the experimental value of $n_c$, although a large experimental value of 
$\Gamma (\ccs)$ will imply that $n_c$ is being underestimated. 
The $(\ccs)$ transition manifests itself as final states containing a $\Ds$, 
$\Xi_c {\bar \Lambda_c}$, or charmonium states. In this section, I will present results which shed some
light on these issues.

\begin{figure}[htb]
\vspace{9pt}
\hspace{0mm}
\vspace{-0.5cm}
{\epsfig{figure=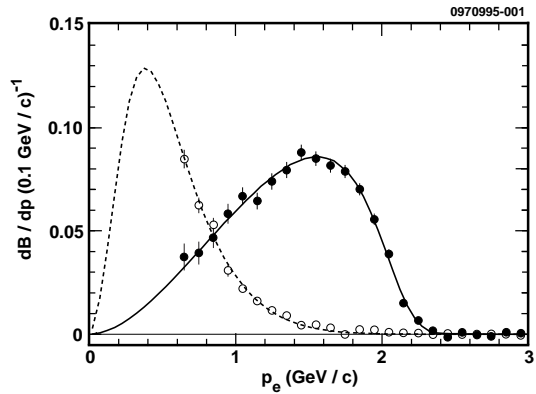, height=6.43 cm}} 
\vspace{-0.5cm}
\caption{Electron Momentum Spectrum.}
\label{fig:double}
\end{figure}

   In fig.~\ref{fig:double}, I show the electron spectrum, $P_e > 0.6$ GeV/c, from B decay, 
where the opposite B has been
tagged with a high momentum lepton (P$_{\rm tag} > 1.5$ GeV/c). Correlating the charge and angle between
the two leptons, we can disentangle the primary lepton spectrum ($b \rightarrow c l \nu$) from the 
secondary spectrum ($b \rightarrow c X, c \rightarrow Y l \nu$). Since we can detect electrons down to
0.6 GeV/c, we are able to probe a larger portion of the momentum spectrum and hence have to rely less
on models to extrapolate down to zero lepton momentum. This analysis yields,
\begin{displaymath}
{\cal B}(B\rightarrow Xl\nu)=(10.49\pm0.17\pm0.43)\%
\end{displaymath}

\subsection{${\bar B} \rightarrow \Ds^+ X$}

        There are two diagrams for producing a $\Ds$ in the final state, (a) $\ccs$: 
external W diagram, 
where $ W~\rightarrow~c{\bar s}$, which hadronizes to form a $\Ds^+$, and, (b) $\cud$: 
internal or external W diagram,
where $ W~\rightarrow~u{\bar d}$, accompanied by $s{\bar s}$ popping. In the second case, the ${\bar c}$
 quark from ${\bar b}$ decay combines with the s quark to form a $\Ds^-$. 

\begin{figure}[htb]
\vspace{9pt}
\hspace{0mm}
\vspace{-0.5cm}
{\epsfig{figure=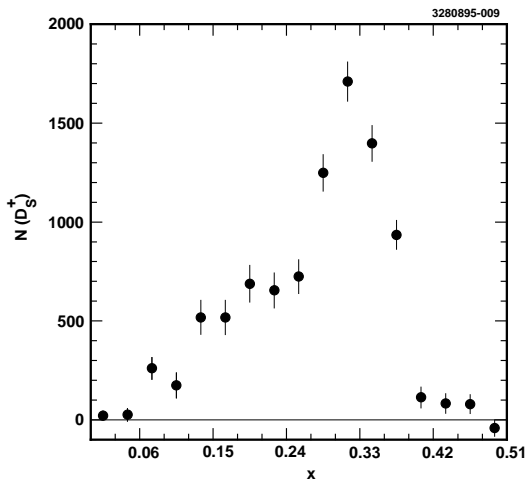, height=6.43 cm}} 
\vspace{-0.5cm}
\caption{$\Ds$ Momentum Spectrum.}
\label{fig:incl_ds}
\end{figure}

    CLEO has measured the inclusive branching fraction\cite{ref:phipi},
\begin{displaymath} 
{\cal B} (B \rightarrow \Ds X) = (11.81 \pm 0.43 \pm 0.94)\%
\end{displaymath} 
This result includes both sources of $\Ds$, as described above. In 
Fig.~\ref{fig:incl_ds}, I show the momentum spectrum of $\Ds$ produced in B decays - the X axis is the
$\Ds$ momentum normalized to the maximum momentum it can have ($[E_{beam}^2 - M_{\Ds}^2]^{1/2}$). The 
data points for $x \ge 0.25$ are due to two-body decays, where the $\Ds$ is produced 
via a $\ccs$ transition, whereas the data points for $x < 0.25$ are either due to 
$\ccs$ where the $\Ds$
is accompanied by more than 1 pion(s) or due to $\cud$, which is always a multi-body final state.

\begin{figure}[htb]
\vspace{9pt}
\hspace{0mm}
\vspace{-0.5cm}
{\epsfig{figure=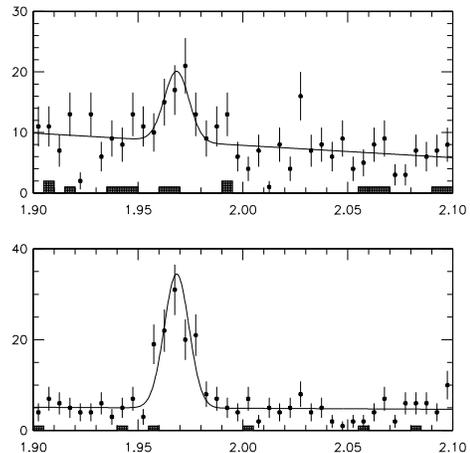, height=6.43 cm}} 
\vspace{-0.5cm}
\caption{$\Ds$ mass in $\Ds - lepton$ combinations.}
\label{fig:ds_lep}
\end{figure}

    To investigate the relative strengths of production mechanism (a), which is a
$\ccs$ transition and, (b), which is a $\cud$ transition, we have used $\Ds - lepton$ correlations, 
where the $\Ds$ 
and the lepton come from different B mesons. The lepton is used to tag the flavour of one B, whereas the
charge of the $\Ds$ is used to tag whether the $\Ds$ is produced by mechanism (a) or (b). Therefore,
$\Ds^- l^-$ combinations imply that the $\Ds$ is produced via (b), whereas $\Ds^+ l^-$ imply that the
$\Ds$ is produced via (a). Fig.~\ref{fig:ds_lep} shows the $\Ds$ mass for the two $\Ds-lepton$ charge
combinations - the $\Ds$ is reconstructed via the $\phi \pi$ decay mode. The raw yield for the 
like-sign and opposite-sign combinations are $34.3\pm 9.1$ and $116.3\pm 15$ events, respectively. After
correcting for backgrounds (shown as black squares) and mixing, 
we find that most of the $\Ds$ mesons are produced via the $\ccs$ transition, 
with at most 31\% produced via  the $\cud$ transition (90\% confidence 
level upper limit).

    At present, this analysis suffers from low statistics, but we hope to complement this
analysis by searching for exclusive decay modes, which will pinpoint more accurately the production 
mechanism for $\Ds$ mesons.

\subsection{B $\rightarrow$ Charmonium}

    This class of decays occurs via an internal W diagram, where $W \rightarrow c{\bar s}$, and the
${\bar c}$ quark produced in the decay of the ${\bar b}$ combines with the c quark to form a charmonium 
state, $J/\psi, \psi', \chi_c, h_c, \eta_c, \psi''$. 
Table~\ref{tab:onia} lists the CLEO measurements of
B decays into charmonium states. A ``direct'' measurement implies that all feed-downs into that final 
state have been removed from the quoted result. Using theoretical estimates for the relative rates of 
$B \rightarrow \chi_{c0}, h_c, \eta_c$, we estimate that the total branching fraction for B to 
charmonium states is $(2.6 \pm 0.3)\%$. Since there are two charm quarks in these states, they enter 
with twice the weight in Table~\ref{tab:nc}.

\begin{table}[hbt]
\setlength{\tabcolsep}{2.0pc}
\caption{Inclusive B decays to Charmonium states.}
\label{tab:onia}
\begin{tabular}{cc}
\hline
  Decay mode &  Rate \\ 
\hline
 ${\bar B} \rightarrow J/\psi X$ (direct) &  $(0.80 \pm 0.08)\% $      \\
 ${\bar B} \rightarrow \psi' X$ (direct) &  $(0.34 \pm 0.05)\% $  \\
 ${\bar B} \rightarrow \chi_{c1} X$ (direct) & $(0.37 \pm 0.07)\% $  \\
 ${\bar B} \rightarrow \chi_{c2} X$ & $(0.25 \pm 0.11)\% $  \\
 ${\bar B} \rightarrow \eta_c X$ & $< 0.9 \% $  \\
\hline
\end{tabular}
\end{table}

\subsection{B $\rightarrow$ baryons}

\begin{figure}[htb]
\vspace{9pt}
\hspace{0mm}
\vspace{-0.5cm}
{\epsfig{figure=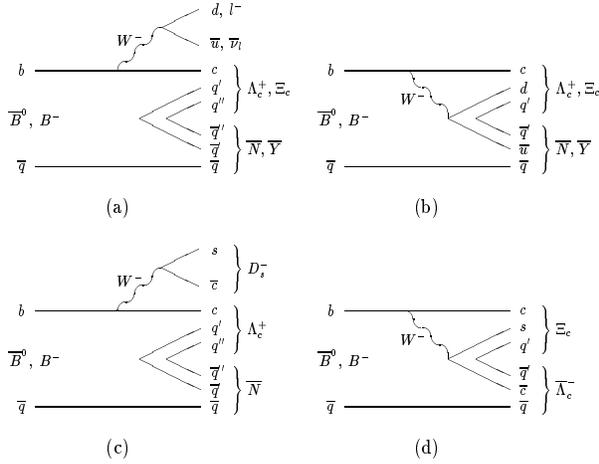, height=6.43 cm}} 
\vspace{-0.5cm}
\caption{Processes for B $\rightarrow$ baryon decays.}
\label{fig:feyn}
\end{figure}

    B $\rightarrow$ baryon decays can be mediated by both $\cud$ and 
$\ccs$ transitions as shown in fig.~\ref{fig:feyn} a-b and c-d, respectively. In this figure, ${\bar N},
{\bar Y}$ represent non-strange (n, p,...) and strange baryons ($\Lambda$,...), respectively. The
external W diagrams ((a), (c)) require two $q{\bar q}$ pairs to be popped from the vacuum, whereas the
internal W diagrams require only one such pair, leading to the possibility that the former class of 
diagrams may not be dominant. In contrast, in B decays to mesons, the external W diagrams are quite
dominant. If the external W diagrams are dominant for B $\rightarrow$ baryons, then $\ccs$ may not 
play a big role here, since they mainly occur in internal W type processes (Fig.~\ref{fig:feyn}c is
phase-space suppressed). In other words, if both external W and $\cud$ are dominant, then one may expect
the ratio ${\bar B} (\Lambda_c {\bar N} X l \nu)/~{\bar B} (\Lambda_c X)\approx 12\%$, 
as is the case for B $\rightarrow$ 
mesons.

    We have studied the importance of external W diagrams, by searching for the decay
$B~\rightarrow~\Lambda_c^+ {\bar N} X e^- \nu$ using $\Lambda_c^+ - e^{\pm}$ correlations, 
where both the 
$\Lambda_c$ and electron come from the same B. Since we have two baryons in the final state, 
the electron
momentum is softer than in the case of B decay to mesons, and we require that it be in the range, 
0.7 GeV/c to 1.5 GeV/c. Opposite-sign combinations, $\Lambda_c^+ e^-$ are due to both signal 
and background events, 
whereas like-sign events $\Lambda_c^+ e^+$ are all background. Background in this case consists 
of picking up the $\Lambda_c^+$ from the decay of one B, and the electron from the other B and also 
due to B mixing. In Table~\ref{tab:sameb} we list the event yields (continuum subtracted) 
and background estimates. 

\begin{table}[hbt]
\setlength{\tabcolsep}{1.0pc}
\caption{$\Lambda_c - e$ combinations from the same B.}
\label{tab:sameb}
\begin{tabular}{ccc}
\hline
 Yields  & $\Lambda_c^+ e^-$ &  $\Lambda_c^+ e^+$ \\ 
\hline
 Raw Yield &  $95 \pm 20$ & $74\pm 16$       \\
 Bkgd estimate &  $57 \pm 13$ &  $87\pm 14$       \\
 Mixing correc. &  $+3 \pm 1$ & $ -3\pm 1$       \\
\hline
 Net Yield      & $35\pm 26$ & $-10\pm 21$ \\
\hline
\end{tabular}
\end{table}


As one
can see, we do not have a statistically significant signal as yet, but with the current data we can set
the following 90\% confidence level upper limit,

\begin{displaymath}
\frac{{\cal B} ({\bar B} \rightarrow \Lambda_c {\bar N} X l \nu)}
{{\cal B} ({\bar B} \rightarrow \Lambda_c X)}~< 6.0\%
\end{displaymath}

   This result implies that the external W diagrams may not be dominant in B $\rightarrow$ baryons, 
because if they were, then the above ratio would be closer to 12\%; thus, we may be able to 
investigate the role of $\ccs$ transitions, which occur mainly in internal W type
processes.

\begin{figure}[htb]
\vspace{9pt}
\hspace{0mm}
\vspace{-0.5cm}
{\epsfig{figure=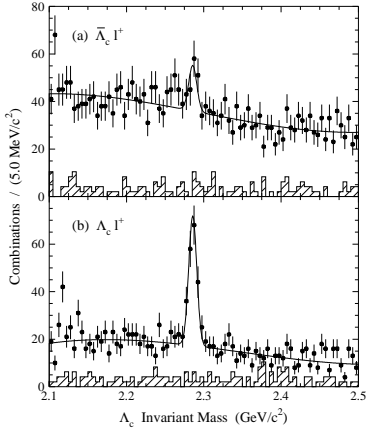, height=6.43 cm}} 
\vspace{-0.5cm}
\caption{$\Lambda_c$ invariant mass for $\Lambda_c e$ combinations from different B's.}
\label{fig:diffb}
\end{figure}

    To investigate the relative strengths of $\cud$ and $\ccs$ transitions, we now look at 
$\Lambda_c - lepton$ correlations, where the two now come from {\bf different B's}. The lepton momentum
is required to be between 1.5 GeV/c and 2.4 GeV/c - this momentum region is relatively free from 
$b \rightarrow c \rightarrow Xl\nu$ contamination. Like sign combinations, $\Lambda_c^+l^+$, arise when
the $\Lambda_c$ is created in a $\cud$ transition (fig.~\ref{fig:feyn}a,b), whereas opposite sign 
combinations, $\Lambda_c^- l^+$, arise when the $\Lambda_c$ is created in a $\ccs$ transition 
(fig.~\ref{fig:feyn}d).  In fig.~\ref{fig:diffb}, I present the $\Lambda_c$ mass for opposite sign and
like sign combinations, respectively, and table~\ref{tab:diffb} lists the raw yields 
(continuum subtracted) and background estimates. The cross-hatched entries in the figure 
are contributions due to continuum background.

\begin{table}[hbt]
\setlength{\tabcolsep}{1.0pc}
\caption{$\Lambda_c - lepton$ combinations from the different B mesons.}
\label{tab:diffb}
\begin{tabular}{ccc}
\hline
 Yields  & $\Lambda_c^+ l^-$ &  $\Lambda_c^+ l^+$ \\ 
         & $\ccs$            &  $\cud$    \\
\hline
 Raw Yield &  $43 \pm 16$ & $141\pm 16$       \\
 Bkgd estimate &  $5 \pm 1.5$ &  $2.1\pm 0.8$       \\
 Mixing correc. &  $-9 \pm 2$ & $ +9 \pm 2$       \\
\hline
 Net Yield      & $29 \pm 19$ & $148\pm 19$ \\
\hline
\end{tabular}
\end{table}


    From these yields, the ratio of the relative strengths of $\ccs$ and $\cud$ 
transitions in $B \rightarrow \Lambda_c$ decays is determined to be,
\begin{displaymath}
 \frac{\Gamma (\ccs)}{{\Gamma}(\cud)}=(20\pm 13\pm 4)\%
\end{displaymath}
nowhere near $2/3$, which is what one may expect if $\Gamma~(\ccs)~\approx~0.3~\Gamma_{total}$ applied
universally to all B decays.
In addition, this result is consistent with the ratio being
$1/3$, which is what one expects from naive phase-space arguments. However, to have a more conclusive
result, we need more data, more techniques of tagging the flavour of one B.

\begin{figure}[htb]
\vspace{9pt}
\hspace{0mm}
\vspace{-0.5cm}
{\epsfig{figure=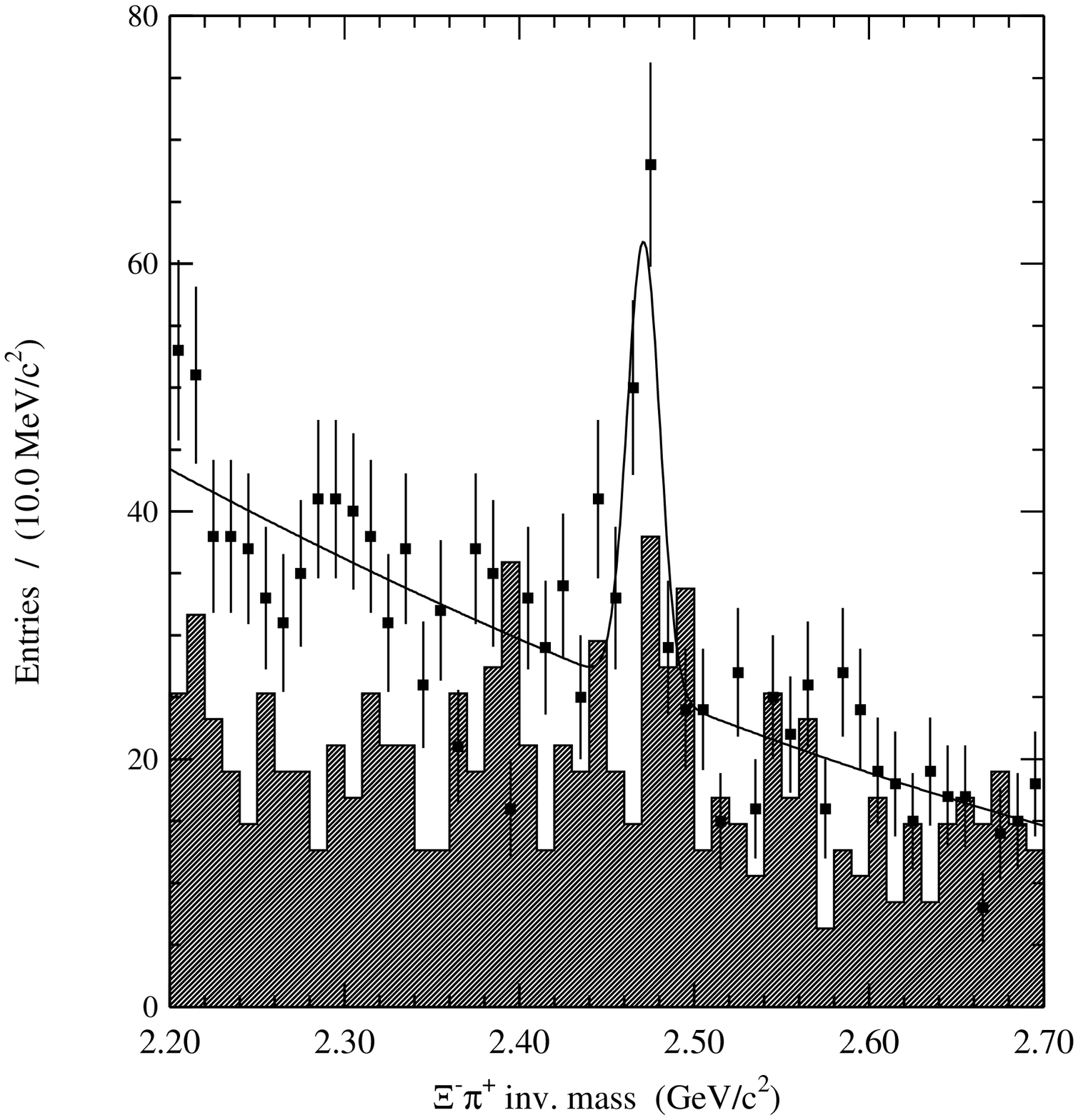, height=6.43 cm}} 
\vspace{-0.5cm}
\caption{$\Xi_c^0$ invariant mass in B decay.}
\label{fig:csc0}
\end{figure}

\begin{figure}[htb]
\vspace{9pt}
\hspace{0mm}
\vspace{-0.5cm}
{\epsfig{figure=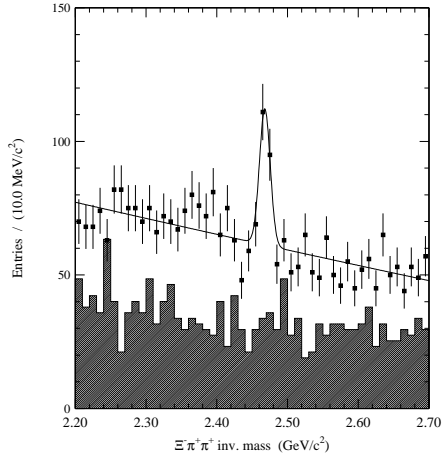, height=6.43 cm}} 
\vspace{-0.5cm}
\caption{$\Xi_c^+$ invariant mass in B decay.}
\label{fig:cscp}
\end{figure}

    $B \rightarrow \Xi_c X$ is another decay mode where one can probe the importance of the $\ccs$ 
transition. This decay mainly occurs via the internal W diagram with the $W\rightarrow u{\bar d}$ 
accompanied by $s{\bar s}$ popping as in fig.~\ref{fig:feyn}b or 
$W\rightarrow c{\bar s}$ accompanied by light quark-pair popping, 
as in fig.~\ref{fig:feyn}d, respectively. There will be also be some contribution due to the external
W diagram as in fig.~\ref{fig:feyn}a.
If $[\ccs/\cud] \approx 1/3$ and the ratio of $s{\bar s}$
to light quark-pair popping is about 0.15, then one could expect the ratio, 
${\cal B}~(B~\rightarrow~\Xi_c~X)~/~{\cal B}~(B~\rightarrow~\Lambda_c~X)~\approx~0.48$. We reconstruct
$\Xi_c^0, \Xi_c^+$ in the $\Xi^- \pi^+, \Xi^- \pi^+ \pi^+$ modes, respectively. The ON (data points) 
and OFF (shaded) resonance
contributions to $\Xi_c^0$ and $\Xi_c^+$ mass distributions are shown in fig.~\ref{fig:csc0} and 
fig.~\ref{fig:cscp}, respectively. We find $59\pm 17$ events for $B \rightarrow \Xi_c^0 X$ and 
$88\pm 20$ events for $B \rightarrow \Xi_c^+ X$.


   To calculate a ratio for inclusive $\Xi_c$ production, we have to estimate the absolute branching
fraction scale for $\Xi_c$ decays. We do this by assuming that the semileptonic widths for all charm
hadrons is the same, and that $\Xi_c \rightarrow \Xi l \nu$ saturates the $\Xi_c$ semileptonic width 
(similarly for $\Lambda_c$). 
This leads to upper limits on the branching fraction of $\Xi_c \rightarrow \Xi X$, and 
$\Lambda_c \rightarrow pK\pi$. I should point out that 
these assumptions are not very reliable, and only serve to make a ``crude'' estimate. Using
CLEO data for the semileptonic data, we get that $B \rightarrow \Xi_c^+ X~=~(2.0 \pm 0.7)\%$,
$B \rightarrow \Xi_c^0 X~=~(2.8 \pm 1.2)\%$, and $B \rightarrow \Lambda_c X~=~(3.1 \pm 1.0)\%$.
Using these estimates, we find that 
$[{\cal B} (B \rightarrow \Xi_c X)/{\cal B} (B \rightarrow \Lambda_c X])~\approx~1.5\pm~0.7$, which is
not terribly conclusive. This result is consistent with a small rate for $\cud$ transitions in 
baryon production, 
which is in sharp disagreement with the result from $\Lambda_c - lepton$ correlations. Most likely, the
branching fraction scale for the charmed baryons is wrong.

\section{Conclusions}

    $\ccs$ transitions do take place, as evidenced by $B \rightarrow \Ds X, \Xi_c {\bar \Lambda_c}X$, 
charmonium states. Our preliminary results indicate that the rate for $\ccs$ is not enough to solve the
${\cal B} (B \rightarrow X l \nu)$ ``problem''. We find this branching fraction to be $(10.49 \pm 0.17 
\pm 0.43)\%$ instead of the expected 12\%, and we also find $n_c$, the number of charm quarks/b quark to
be $1.15 \pm 0.044$ instead of 1.3.

   Lack of time prevents me from presenting other results, but I will briefly point out some of them.
\begin{itemize}
\item We have made the first unambiguous measurement of $\Ds$ semileptonic decays 
to $\eta, \eta'$ final states.
\begin{displaymath}
\frac{{\cal B} (\Ds \rightarrow \eta l \nu)} {{\cal B} 
(\Ds \rightarrow \phi l \nu)}~=~1.24 \pm 0.12 \pm 0.15
\end{displaymath}
\begin{displaymath}
\frac{{\cal B} (\Ds \rightarrow \eta' l \nu)} {{\cal B} 
(\Ds \rightarrow \phi l \nu)}~=~0.43 \pm 0.11 \pm 0.07
\end{displaymath}
    The ratio of the vector to pseduoscalar final states in $\Ds$ semileptonic decays is about the same
as one finds in non-strange D semileptonic decays ($\approx 0.6$). In the past, most theoretical models
predicted this ratio to be 1. 

\item We have made the first measurement of exclusive $b \rightarrow u$ decays,
\begin{displaymath}
{\cal B}(B^0\rightarrow\pi^+l^-\nu)~=~(1.34\pm 0.35\pm 0.28)\times 10^{-4}
\end{displaymath}
\begin{displaymath}
{\cal B} (B^0\rightarrow\rho^+l^-\nu)~=~(2.28\pm 0.36\pm 0.59^{+0.00}_{-0.46})\times 10^{-4}
\end{displaymath}
These branching fractions have been obtained using isospin constraints between the final states 
$\pi^0l\nu$ and $\pi^+l\nu$, and between $\rho^0l\nu, \rho^+l\nu$ and $\omega l\nu$. 
The ISGW model was used to determine efficiencies, etc.
\end{itemize}

   Currently, we are processing more data which has already been collected. To further increase the 
luminosity of CESR and the capabilities of the CLEO detector various upgrades are 
underway. A new silicon vertex detector is being installed in CLEO and in 3-4 years we are planning to
significantly improve particle identification in CLEO \cite{ref:galik}. With these improvements, we 
expect to be doing exciting physics in the future.

\section{Acknowledgements}

    I would like to thank my colleagues on CLEO for explaining to me the details of their analyses. I 
also thank Scott Menary and Isi Dunietz for their comments. This research was funded by the U.S. 
Department of Energy, National Science Foundation and Vanderbilt University.



\begin{thebibliography}{99}
\bibitem{ref:note} The values for B $\rightarrow$ baryons are somewhat uncertain due to the lack of
knowledge of the branching ratio scale for charm baryon decays.
\bibitem{ref:phipi} We have taken ${\cal B} (\Ds \rightarrow \phi \pi) = (3.7 \pm 0.9)\%$ for this 
result.
\end{thebibliography}
\end{document}